\documentclass[pra, 12 pt]{revtex4}
\usepackage[english]{babel}
\usepackage{amsmath}
\usepackage{epsfig}

\begin{document}
\title{Diffusion in a Time-dependent External Field}
\author{S.A. Trigger $^1$, G.J.F. van Heijst $^2$, O.F. Petrov $^1$ and P.P.J.M. Schram $^2$}
\address{$^1$ Joint\, Institute\, for\, High\,
Temperatures, Russian\, Academy\, of\, Sciences, 13/19, Izhorskaia
Str., Moscow\, 127412, Russia;\\
email:\,strig@gmx.net \\
 $^2$ Eindhoven  University of Technology, P.O. Box 513, MB
5600 Eindhoven, The Netherlands}

\begin{abstract}
The problem of diffusion in a time-dependent (and generally
inhomogeneous) external field is considered on the basis of a
generalized master equation with two times, introduced in [1,2].
We consider the case of the quasi Fokker-Planck approximation,
when the probability transition function for diffusion
(PTD-function) does not possess a long tail in coordinate space
and can be expanded as a function of instantaneous displacements.
The more complicated case of long tails in the PTD will be
discussed separately. We also discuss diffusion on the basis of
hydrodynamic and kinetic equations and show the validity of the
phenomenological approach. A new type of "collision" integral is
introduced for the description of diffusion in a system of
particles, which can transfer from a moving state to the rest
state (with some waiting time distribution). The solution of the
appropriate kinetic equation in the external field also confirms
the phenomenological approach of the generalized master equation.

\end{abstract}.

\maketitle

\section{Introduction}

Models of continuous time random walks (CTRW) [3], for objects
that may jump from one point to another in a generally
inhomogeneous medium and which may stay in these points for some
time before the next usually stochastic jump, are important for
the solution of many physical, chemical and biological problems.
Recently these models have been applied also in economics and in
social sciences (see, e.g., [4-6]). Usually the stochastic motion
of the particles leads to a second moment of the density
distribution that is linear in time $<r^2(t)>\sim t$. Such type of
diffusion processes play a crucial role in plasmas, including
dusty plasma [7], in nuclear physics [8], in neutral systems in
various phases [9] and in many other problems. However, in many
systems the deviation from the linear time dependence of the mean
square displacement have been experimentally observed, in
particular, under essentially non-equilibrium conditions or for
some disordered systems. The average square separation of a pair
of particles passively moving in a turbulent flow grows, according
to Richardson's law, with the third power of time [10]. For
diffusion typical for glasses and related complex systems [11] the
observed time dependence is slower than linear. These two types of
anomalous diffusion obviously are characterized as superdiffusion
and subdiffusion.

The generalized master equation for the density evolution, which
describes the various cases of normal and anomalous diffusion has
been formulated in [1,2] by introduction of the specific kernel
function (PTD) $W({\bf r, \bf r'}, \tau, t-\tau)$ depending on two
times, which connects in a linear way the density distributions
$f$ of the stochastic objects (or particles) for the points ${\bf
r'}$ at moment $\tau$ and ${\bf r}$ at moment $t$. The approach
suggested in [1,2] clearly demonstrates the relation between the
integral approach and the fractional differentiation method [12]
and permits to extend (in comparison with the fractional
differentiation method) the class of sub- and superdiffusion
processes, which can be successfuly described. On this basis
different examples of superdiffusive and subdiffusive processes
were considered in [2] for the various kernels $W$ and the
mean-squared displacements have been calculated. The idea of the
generalized master equation with two times [1,2] for diffusion in
coordinate space has been recently used in [13] for the
calculation of average displacements in the case of a
time-dependent homogeneous external field. In [13] the jumps of
the particles are assumed to be instantaneous, all particles are
practically trapped and the electric field does not act on the
waiting probability, which is independent of the external
(electric) field. In this conditions the characteristic time scale
of the external field has to be large (in comparison with the
other time scales of the problem) and the probability of jumps is
connected locally in time with the external field. As the result,
in the diffusion equation the external field is placed outside of
the integral on time.

It should be noted, however, that in general case of the problem
of diffusion in a time-dependent external field the force is
placed under the integral over $\tau$ (see the
semi-phenomenological consideration in [14] and Eqs.~(\ref{F2b}),
(\ref{F2c}) below).

The general phenomenological approach to this problem has been
formulated in [14].

This paper is motivated by the necessity to describe in more
detail the influence of time- and space dependent external fields
on the continuous-time random walks. The equation formulated in
[1,2] is appropriate for this purpose and offers the opportunity
for consideration of CTRW for both cases: long-tail space behavior
of the PTD function, as well as for the fast decay of PTD function
in coordinate space, when the Fokker-Planck type expansion is
applicable. For simplicity we consider in this paper only the last
case.

\section{Generalized Master equation}

Let us start from the generalized master equation with two times
[1,2]:
\begin{equation}
f({\bf r},t) = f({\bf r},t=0)+ \int_0^t d\tau \int d{\bf r'}
\left\{W ({\bf r, r'},\tau, t-\tau) f({\bf r',\tau}) - W ({\bf r',
r},\tau, t-\tau) f({\bf r},\tau) \right\}. \label{F1}
\end{equation}
Equation (\ref{F1}) can be represented in an equivalent form, more
similar to the structure of the Fokker-Planck equation, where the
initial condition is absent:
\begin{equation}
\frac{\partial f({\bf r},t)}{\partial t} = \frac{d}{dt} \int_0^t
d\tau \int d{\bf r'} \left\{W ({\bf r, r'},\tau, t-\tau) f({\bf
r',\tau}) - W ({\bf r', r},\tau, t-\tau) f({\bf r},\tau) \right\}.
\label{F2}
\end{equation}
or
\begin{equation}
\frac{\partial f({\bf r},t)}{\partial t} = \int_0^t d\tau \int
d{\bf r'} \left\{P ({\bf r, r'},\tau, t-\tau) f({\bf r',\tau}) - P
({\bf r', r},\tau, t-\tau) f({\bf r},\tau) \right\}, \label{F3}
\end{equation}
where the PTD-function $P ({\bf r, r'},\tau, t-\tau)$ is given by:
\begin{equation}
P ({\bf r, r'},\tau, t-\tau)\equiv 2W ({\bf r', r},\tau,
t-\tau)\delta (t-\tau)+ \frac{\partial}{\partial t} W ({\bf r',
r},\tau, t-\tau)\label{F4}
\end{equation}
Apparently,  different - but equivalent - forms of the master
equation exist with different kernels, although connected
analytically. The form ~(\ref{F3}) is more similar to the form
introduced first in the papers [14-16], where memory effects have
been considered in a very general form on the basis of a master
equation with one time argument $t-\tau$, which describes the
retardation (or memory) effects. It should be stressed, that in
[16], in particular, the straightforward connection of the
generalized master equation (GME) with the usual CTRW model has
been established. In the framework of the specific multiplicative
regime of the function $P ({\bf r, r'}, t-\tau)=\tilde P ({\bf r,
r'})\zeta(t-\tau)$ the dependence of $P({\bf r, r'})$ and
$\zeta(t-\tau)$ on the waiting time distribution and the jump
length distribution is quite clear (see Eqs.~(9),(10) in [15]).
The same applies to the function $W$, which is connected with $P$
by Eq.~(\ref{F4}). Similar problems for the kernel, depending on
one time variable, have been discussed in [17]. In our further
consideration we will derive the memory function as a function of
the waiting time following the same line as in the papers [14-16]
and we find the additional retardation function, which is the
retardation of the mobility under the action of an external force
(physically similar to dispersion of conductivity after
Fourier-transformation in time). A description of this new
retardation function depends on the specific model for the
mobility and this will be considered in a separate paper.
 The argument $t-\tau$ describes the retardation (or memory)
effects, which can be connected in the particular case of
multiplicative PTD function $W ({\bf r, r'},\tau, t-\tau)\equiv
\tilde W ({\bf r, r'},\tau) \chi(t-\tau)$ with, for example, the
probability for particles to stay during some time at a fixed
position before moving to the next point. An equation with
retardation, with the $W$ function depending only on one time
argument $t-\tau$, has been suggested first in [15] and applied in
[16] to the case of the multiplicative representation of the PTD
function. In general $W$ is not a multiplicative function in the
sense mentioned above and, what is more important, is a function
of two times $t$ and $t-\tau$ [1]. It should be mentioned that the
closed form of the equation for the density distribution is an
approximation. In some cases the exact solution for density
distribution can be found (see e.g. [16]-[19]), when a closed
equation for the density distribution does not exist or gives a
too rough approximate result. Nevertheless, in many practical
situations Eqs.~(\ref{F1}) or (\ref{F3}) are sufficiently exact
and permit to describe various experimental data.

Let us consider the role of appearance of the two time arguments
in the generalized master equation Eq.~(\ref{F1}) for the case of
a time-dependent external force ${\bf F}({\bf r},t)$. To simplify
the consideration we can investigate the case of fast decay of the
kernel $W ({\bf r, r'},\tau, t-\tau)\equiv W ({\bf u, r},\tau,
t-\tau)$ as a function of ${\bf u=r-r'}$, when an expansion in the
spirit of Fokker-Planck can be applied. In this case
Eq.~(\ref{F1}) takes the form [1,2]:
\begin{eqnarray}
f({\bf r},t) = f({\bf r},t=0)+ \int_0^t d\tau {\partial \over
{\partial r_\alpha}} \left[A_\alpha ({\bf r},\tau, t-\tau) f({\bf
r},\tau) + {\partial \over {\partial r_\beta}} \left(
B_{\alpha\beta}({\bf r},\tau, t-\tau) f({\bf r},\tau)
\right)\right], \label{F5}
\end{eqnarray}
where the functions $A_\alpha ({\bf r},\tau, t-\tau)$ and
$B_{\alpha\beta}({\bf r},\tau, t-\tau) f_g({\bf r},\tau)$ are the
functionals of the PTD function (the indices are equal
$\alpha,\beta=x_s$ in s-dimensional coordinate space):
\begin{eqnarray}
A_\alpha({\bf r},\tau, t-\tau) = \int d^s u u_\alpha W({\bf u,
r},\tau, t-\tau) \label{F6}
\end{eqnarray}
and
\begin{eqnarray}
B_{\alpha\beta}({\bf r},\tau, t-\tau)= \frac{1}{2}\int d^s u \,
u_\alpha u_\beta W({\bf u, r},\tau, t-\tau). \label{F7}
\end{eqnarray}
Eq.~(\ref{F5}) can be rewritten naturally in a form similar to
Eq.~(\ref{F2}), but now for the Fokker-Planck type approximation:
\begin{eqnarray}
\frac{\partial f({\bf r},t)}{\partial t} = \frac{d}{dt} \int_0^t
d\tau {\partial \over {\partial r_\alpha}} \left[A_\alpha ({\bf
r},\tau, t-\tau) f({\bf r},\tau) + {\partial \over {\partial
r_\beta}} \left( B_{\alpha\beta}({\bf r},\tau, t-\tau) f({\bf
r},\tau) \right)\right], \label{F2a}
\end{eqnarray}
We suggest, that the PTD function is independent of $f({\bf
r},t)$, therefore the problem is linear.

\section{Influence of the external fields}

One of the main sources of inhomogeneity is an external field,
which also provides the prescribed dependence of the PTD function
on $\tau$. Other words we can suggest, in the particular case
considered, that the dependence of $W({\bf u, r},\tau, t-\tau)$ on
the arguments ${\bf r},\tau$ is connected with a functional
dependence on the external field:
\begin{equation}
W({\bf u, r},\tau, t-\tau)=W({\bf u}, t-\tau; {\bf F}({\bf
r},\tau)).\label{F8}
\end{equation}
If an external field is absent the PTD function is a function of
the modulus ${\bf u}\equiv u$, which implies that $A_\alpha=0$ and
$B=\delta_{\alpha\beta}B_0(t-\tau)$ with:
\begin{equation}
B_0(t-\tau)= \frac{1}{2s}\int d^s u \, u^2
W_0(u,t-\tau).\label{F7a}
\end{equation}

For relatively weak external fields the functional (\ref{F8}) can
be linearized as:
\begin{equation}
W({\bf u}, t-\tau; {\bf F}({\bf r},\tau))=W_0(u, t-\tau)+ W_1(u,
t-\tau)({\bf u} \cdot {\bf F}({\bf r},\tau)). \label{F9}
\end{equation}
The functions $W_0(u, t-\tau)$ and $W_1(u, t-\tau)$ are equal to
$W({\bf u}, t-\tau; {\bf F}=0)$ and the functional derivative
$\delta W({\bf u}, t-\tau; {\bf F}({\bf r},\tau))/\delta({\bf u}
\cdot {\bf F}({\bf r},\tau))_{|\textbf{F}=0}$ respectively. Then
the functions $A_\alpha$ and $B_{\alpha\beta}$ take the form
\begin{eqnarray}
A_\alpha({\bf r},\tau, t-\tau) =\frac{1}{s}{\bf F}_\alpha ({\bf
r},\tau)\int d^s u u^2 W_1(u, t-\tau)\equiv {\bf F}_\alpha ({\bf
r},\tau)L(t-\tau),\label{F10}
\end{eqnarray}
where $L(t-\tau)$ is given by
\begin{eqnarray}
L(t-\tau) =\frac{1}{s}\int d^s u u^2 W_1(u, t-\tau).\label{F10a}
\end{eqnarray}
and
\begin{eqnarray}
B_{\alpha\beta}({\bf r},\tau, t-\tau)=
\delta_{\alpha\beta}B_0(t-\tau). \label{F11}
\end{eqnarray}
The generalized diffusion equation Eq.~(\ref{F2a}) takes the form
\begin{eqnarray}
\frac{\partial f({\bf r},t)}{\partial t} = \frac{d}{dt} \int_0^t
d\tau \left[L(t-\tau)\nabla({\bf F} ({\bf r},\tau) f({\bf
r},\tau)) + B_0(t-\tau){\Delta} f({\bf r},\tau)\right].
\label{F2b}
\end{eqnarray}
In general this equation contains two different functions $B_0$
and $L$ depending on the argument $t-\tau$. For the case of a
time-independent inhomogeneous one-dimensional external field and
in the particular case of the kernel dependence on time
$L(t-\tau)\sim(t-\tau)^{\gamma-1}$ and
$B_0(t-\tau)\sim(t-\tau)^{\gamma-1}$ ($0<\gamma<1$) we arrive at
the result, obtained in [20],[21] for the fractional Fokker-Planck
equation. This kind of time dependence for the kernel is typical
for the subdiffusion processes.

The time-dependent mobility for the diffusion process (in the
particular case of exponentially oscillating time-dependent
external field and a time-independent diffusion coefficient) has
been introduced in [22].

If the functional $W({\bf u}, t-\tau; {\bf F}({\bf r},\tau))$ is
multiplicative, namely, $W({\bf u}, t-\tau; {\bf F}({\bf
r},\tau))=\tilde W({\bf u}; {\bf F}({\bf r},\tau))\chi(t-\tau)$
Eq.~(\ref{F2b}) can be simplified to:
\begin{eqnarray}
\frac{\partial f({\bf r},t)}{\partial t} = \frac{d}{dt} \int_0^t
d\tau \chi(t-\tau)\left[D {\Delta} f({\bf r},\tau)-b\nabla({\bf F}
({\bf r},\tau) f({\bf r},\tau))\right], \label{F2c}
\end{eqnarray}
Here $b$ and $D$ are constants, determined by the relations:
\begin{eqnarray}
b =-\frac{1}{s}\int d^s u u^2 \tilde W_1(u)\label{F10b}
\end{eqnarray}
with $\tilde W_1(u)=\delta \tilde W({\bf u}; {\bf F}({\bf
r},\tau))/\delta({\bf u} \cdot {\bf F}({\bf
r},\tau))_{|\textbf{F}=0}$ and
\begin{eqnarray}
D =\frac{1}{2s}\int d^s u u^2 \tilde W_0(u).\label{F10a}
\end{eqnarray}
As is easy to see for the external field ${\bf F}({\bf r},\tau)$,
which change slow in time (comparing with other characteristic
time scales of the problem, e.g., with the time scale of the
retardation function $\chi(t-\tau)$) Eq.~(\ref{F2c}) coincides for
one-dimensional case with the diffusion equation in [13].

The physical meaning of the multiplicative structure of the
functional $W$ is that the independence of the time delay of the
random walkers is independent of the external field. The
dimensionless function $\chi(t)$ in this simple case is associated
with the hopping-distribution function $\psi(t)=\lambda
\psi^\ast(\lambda t)$ introduced in the master equation by Scher
and Montroll [15], with $\lambda\equiv 1/\tau_0$ ($\tau_0$ is the
characteristic waiting time for the hopping-distribution). Laplace
transformations of these functions $\chi(z)$ and $\psi^\ast(z)$
relate them as follows
\begin{eqnarray}
\chi(z)=\frac{\psi^\ast(z)}{1-\psi^\ast(z)}.\label{F12}
\end{eqnarray}
For an exponential hopping-time distribution $\psi(t)=\lambda
exp(-\lambda t)$, where $\lambda\equiv1/\tau_0$  we have
$\psi^\ast(z)=1/(1+z)$, $\chi(z)=1/z$ and
$\chi(t)\equiv\chi(\lambda t)=1$. In this case Eq.~(\ref{F2c})
reduces to the usual diffusion equation in an external field with
diffusion coefficient $D$ and mobility $b$:
\begin{eqnarray}
\frac{\partial f({\bf r},t)}{\partial t} = D {\Delta} f({\bf
r},t)-b\nabla\left({\bf F} ({\bf r},t) f({\bf r},t)\right).
\label{F13}
\end{eqnarray}

\section{Hydrodynamic approach}

In order to better understand the situation on the basis of a
non-phenomenological approach, let us consider the charged
particles with an inhomogeneous density in the external electrical
field in the hydrodynamic approximation. The equation for the
density $n(x,t)$ reads
\begin{eqnarray}
\frac{\partial}{\partial t}n(x,t)+ div{\bf j(x,t)}=0, \label{A1}
\end{eqnarray}
where ${\bf j(x,t)}=n(x,t){\bf v(x,t)}$ and ${\bf v(x,t)}$ is the
hydrodynamic velocity. In the hydrodynamic approximation, when the
charged particles particles (with charge $e$ and mass $m$) move in
the medium under the action of an external time-dependent
electrical field ${\bf E(x,t)}$ the equation of motion has (for
constant temperature $T$) the form
\begin{eqnarray}
\frac{\partial}{\partial t}[{n(x,t) v_i(x,t)}]+ \nabla_k [n(x,t)
v_i(x,t) v_k(x,t)]\\ \nonumber =-\frac{T}{m}\nabla_i n(x,t)+
\frac{e}{m} E_i(x,t)n(x,t)-\nu n(x,t) v_i(x,t). \label{A2}
\end{eqnarray}
Here $\nu$ is the effective frequency of collision with the
particles of the thermostat. In the linear by ${\bf v}$
approximation the solution of Eq.~(\ref{A2}) gives the closed
expression for the flux ${\bf j}$ via the density $n(x,t)$. This
solution for time-independent $\nu$ has the form
\begin{eqnarray}
{\bf j(x,t)}= \int_{-\infty}^t dt'exp\,[-\nu(t-t')]
\left\{\frac{e}{m}[n(x,t'){\bf E(x,t')}] -\frac{T}{m}\nabla
n(x,t') \right\}. \label{A3}
\end{eqnarray}
Inserting this value of ${\bf j(x,t)}$ in Eq.~(\ref{A1}) leads to
the diffusion equation
\begin{eqnarray}
\frac{\partial n(x,t)}{\partial t}=- \int_{-\infty}^t dt'\left
\{D(t-t')\triangle n(x,t')- e\mu(t-t') \nabla [n(x,t'){\bf
E}(x,t')] \right\}, \label{A4}
\end{eqnarray}
where in the case considered the "effective diffusion function"
and "effective mobility function" are given by $D(t)\equiv
T\,exp(-\nu t)/m$ and $\mu(t)\equiv exp(-\nu t)/m$, respectively.
If the functions ${\bf E(x,t)}$ and $n(x,t)$ change in time very
slowly (the characteristic time for its change $\tau \gg 1/\nu$)
Eq.~(\ref{A4}) reduces to the standard form of the diffusion
equation
\begin{eqnarray}
\frac{\partial n(x,t)}{\partial t}=D_0 \triangle n(x,t)-e\mu_0
\nabla [n(x,t){\bf E(x,t)}]. \label{A5}
\end{eqnarray}
Here we introduced the notations $D_0=T/m\nu$ for the diffusion
coefficient and $\mu_0=1/m\nu$ for the mobility coefficient.

Equation (\ref{A4}) represents a particular case (in hydrodynamic
approximation) of the general relations between the fluxes and
acting thermodynamical and the external forces. Of cause, the time
integration in Eq.~(\ref{A4}) can be considered in the normal
hydrodynamical conditions as an excess of accuracy due to the
inequality $\tau \gg 1/\nu$. For us, however, the most important
result is the general structure of Eq.~(\ref{A4}), which
demonstrates that the time integral includes the electrical field
$\bf {E(x,t)}$. The structure of Eq.~(\ref{A4}) confirms the
result of our consideration on the basis of the generalized master
equation for diffusion [14], where the time dependent electric
field included in the time integration.

Since the equilibrium density in the external time-independent
potential $\varphi(x)$ has a form of the Boltzmann distribution
$n(x)~\sim exp\,[-\varphi(x)/T]$, the diffusion and mobility
coefficients satisfy the Einstein relation $D_0=\mu_0 T$. In the
considered case the same statement is valid also for the effective
diffusion and mobility functions $D(t)$ and $\mu(t)$, namely
$D(t)=T \mu(t)$. The general structure of the diffusion equation
(\ref{A4}) is similar to the phenomenological Eq.~(\ref{F2c})
(with the appropriate renormalization of the kernel, which
eliminates the external derivative of the time integral).

\section{Kinetic approach}

Let us start with the kinetic equation for the distribution
function in an electric field
\begin{eqnarray}
\frac{\partial f(p,x,t)}{\partial t}+v \frac{\partial
f(p,x,t)}{\partial x}+e E(x,t)\frac{\partial f(p,x,t)}{\partial
p}= I_{st}(p,x,t). \label{F26}
\end{eqnarray}
Here $I_{st}$ is some kind of "collision integral", which can
describe in general, as we show below, not only real collisions of
particles, but also (for the appropriate problems, e.g. moving of
the alive objects) the more complicated processes, as the
displacements with some pauses, etc.

For simplicity we consider the one-dimensional case $s$=1, but the
generalization for the cases $s$=2,3 is trivial. The distribution
function $f(p,x,t)$ is normalized to the density $n(x,t)$
\begin{eqnarray}
\int dp f(p,x,t)=n(x,t). \label{F26a}
\end{eqnarray}
For the case when the collision integral conserves the total
number of particles, i.e.
\begin{eqnarray}
\int dp I_{st}(p,x,t)=0, \label{F27}
\end{eqnarray}
integration by $p$ leads to the continuity equation
\begin{eqnarray}
\frac{\partial n(x,t)}{\partial t}+ div j(x,t)=0. \label{F27}
\end{eqnarray}

To calculate the flux $j(x,t)$ let us use the Fokker-Planck
approximation for the collision integral $I_{st}(p,x,t)$ and
rewrite for this case Eq.~(\ref{F26}) in the form
\begin{eqnarray}
\frac{\partial f(p,x,t)}{\partial t}+v \frac{\partial
f(p,x,t)}{\partial x}+e E(x,t)\frac{\partial f(p,x,t)}{\partial
p}=\frac{\partial}{\partial p}\left(\beta p f(p,x,t)+m^2 \tilde
D\frac{\partial f(p,x,t)}{\partial p}\right) \label{F28}
\end{eqnarray}
We suggest that the friction $\beta$ and the diffusion $\tilde D$
coefficients in velocity space are the constants, which satisfies
the Einstein relation $\beta T=m \tilde D$. Integrating
Eq.~(\ref{F28}) by $p$ leads to the expression
\begin{eqnarray}
\frac{\partial j(x,t)}{\partial t}+\frac{\partial}{dx}\left[\int
dp v^2 f(p,x,t)\right]-\frac{e}{m} E(x,t)n(x,t) = -\beta j(x,t)
\label{F29}
\end{eqnarray}
If we assume that $f(p,x,t)$ has the quasi-equilibrium form
$f(p,x,t)=n(x,t)f_0(p)$, then we arrive at the following solution
of Eq.~(\ref{F29}) similar to (\ref{A3})
\begin{eqnarray}
j(x,t)= \int_{-\infty}^t dt'exp\,[-\beta(t-t')]
\left\{\frac{e}{m}[n(x,t') E(x,t')] -<v^2> \nabla n(x,t')
\right\},\label{A30}
\end{eqnarray}
where for the Maxwellian distribution $f_0(p)=F_M(p)$ in
one-dimensional ($s=1$) case  $<v^2>=T/m$. In this case the
diffusion equation is equivalent to Eq.~(\ref{A4}) obtained in the
hydrodynamic approach, but with the change $\nu\rightarrow \beta$
in the functions $D(t)$, $\mu(t)$, as well as in the coefficients
$D_0$ and $\mu_0$. The function D(t) is naturally connected with
the time dependent conductivity $\sigma(t)=e^2 n_0 \mu(t)$, where
$n_0$ is the average density of the particles. In the simple case
considered the respective frequency-dependent conductivity
$\sigma(\omega)$ is
\begin{eqnarray}
\sigma(\omega)= \frac{ie^2 n_0}{m (\omega+i\nu)}. \label{A30a}
\end{eqnarray}

Let us now consider the alternative case of the kinetic equation
(\ref{F26}), when the collisions are negligible
($I_{st}=-\varepsilon f(p,x,t)$ with $\varepsilon\rightarrow 0$).
We also suppose that the electric field is weak and can be
considered as a perturbation. To find the evolution of the density
we split the distribution function in two parts:
$f(p,x,t)=f_0(p,x,t)+f_1(p,x,t)$, where the perturbation $f_1$ is
proportional to the electric field $E(x,t)$. The respective
kinetic equations are
\begin{eqnarray}
\frac{\partial f_0(p,x,t)}{\partial t}+v \frac{\partial
f_0(p,x,t)}{\partial x}=0; \; \,f_0=f_0(x-vt,p) \label{F31}
\end{eqnarray}
\begin{eqnarray}
\frac{\partial f_1(p,x,t)}{\partial t}+v \frac{\partial
f_1(p,x,t)}{\partial x}+e E(x,t)\frac{\partial
f_0(p,x,t)}{\partial p}=-\varepsilon f_1(p,x,t)\label{F32}
\end{eqnarray}
The continuity equations follow from Eqs.~(\ref{F31}),
(\ref{F32}):
\begin{eqnarray}
\frac{\partial n_0(x,t)}{\partial t}+ div j_0(p,x,t)=0,
\label{F31a}
\end{eqnarray}
where $j_0(x,t)$ describes the flux without the electrical field,
and
\begin{eqnarray}
\frac{\partial n_1(x,t)}{\partial t}+div j_1(x,t)=0, \label{F32a}
\end{eqnarray}
where $j_1(x,t)$ describes the perturbation of the flux in the
lowest order of the electric field.

The solution of Eq.~(\ref{F32}) reads
\begin{eqnarray}
f_1(p,x,t)=-e \int_{-\infty}^t dt'exp\,[-\varepsilon (t-t')]
\frac{\partial f_0(x-v t,p)}{\partial p} E(x-v(t-t'),t').
\label{F33}
\end{eqnarray}

Now we can calculate $j(x,t)=j_0(x,t)+j_1(x,t)$:
\begin{eqnarray}
j_0 (x,t)= \int dp v f_0(x-vt, p)\label{F34}
\end{eqnarray}
\begin{eqnarray}
j_1 (x,t)= \int dp v f_1(p,x,t)= -e \int_{-\infty}^t
dt'exp\,[-\varepsilon (t-t')]\times \\ \nonumber  \int dp v
\frac{\partial f_0(x-vt, p)}{\partial p}
E(x-v(t-t'),t')\label{F35}
\end{eqnarray}
The latter equation can be rewritten as
\begin{eqnarray}
j_1 (x,t)=  -e \int_{-\infty}^t dt'exp\,[-\varepsilon (t-t')]\times \\
\nonumber \int d x'\int dp v \frac{\partial
f_0(p,x'-vt')}{\partial p} \delta(x-x'-v(t-t')) E(x',t')\nonumber\\
\equiv \int_{-\infty}^t dt'exp\,[-\varepsilon (t-t')]\int d
x'\pi(x,x',t,t')E(x',t'). \label{F36}
\end{eqnarray}
In equation Eq.~(\ref{F36}) the function $\pi(x,x',t,t')$ is equal
to
\begin{eqnarray}
\pi (x,x',t,t')=  -e \int dp v \frac{\partial
f_0(x-vt,p)}{\partial p} \delta(x-x'-v(t-t')), \label{F36a}
\end{eqnarray}
in which $f_0(x-vt,p)$ can also be written as $f_0(x'-vt',p)$. The
function $\pi (x,x',t,t')$ takes into account the processes of
space and time dispersion for the inhomogeneous and time-dependent
distribution $f_0=f_0(x-vt,p)$.

Let us choose the distribution function $f_0$ in the natural form
$f_0(x-vt,p)=n_0(x-vt)f_0(p)$. Then finally we arrive at the
expressions for the fluxes $j_0(x,t)$ and $j_1(x,t)$:
\begin{eqnarray}
j_0 (x,t)= \int dp v n_0(x-vt)f_0(p)\label{F37}
\end{eqnarray}
\begin{eqnarray}
j_1 (x,t)=-e \int_{-\infty}^t dt'exp\,[-\varepsilon (t-t')]\int d
x'\int dp v \frac{\partial [f_0(p)n_0(x-vt)]}{\partial p}
\delta(x-x'-v(t-t'))E(x',t'), \label{F38}
\end{eqnarray}

The expression for $\pi(x,x',t,t')$ can be rewritten in the form
\begin{eqnarray}
\pi (x,x',t,t')= - e \int dp v \left[n_0(x-vt)\frac{\partial
f_0(p)}{\partial p} - \frac{t}{m} f_0(p)\nabla n_0(x-vt)
\right]\delta(x-x'-v(t-t'))\label{F39}
\end{eqnarray}
Here and in what follows the operator $\nabla_x$ acts only on the
function $n_0$ placed behind it. After integration by $v$ we find
\begin{eqnarray}
\pi (x,x',t,t')= - e m \frac{x-x'}{(t-t')^2}\left\{
n_0[(x't-xt')/(t-t')]\frac{\partial f_0(p)}{\partial
p}|_{p=m(x-x')/(t-t')}- \right.\nonumber\\
\left. \frac{t}{m} \nabla n_0[(x't-xt')/(t-t')]
f_0(p)|_{p=m(x-x')/(t-t')} \right\}. \label{F40}
\end{eqnarray}

If $E(x,t)$ is an oscillating function proportional to $sin
(\omega t)$ or $cos (\omega t)$ or a function damping in time, the
argument $x-vt$ under the integral in Eq.~(\ref{F38}) equals to
$(x't-xt')/(t-t')$. The expression of the particle density
$n_0(x-vt)$ (due to the presence of $\delta$-function) in the
limit of large $t$ can be taken equal to $x'$. In this case the
function $\pi$ can then in good approximation be written in the
form
\begin{eqnarray}
\pi (x,x',t,t')= - e m \frac{x-x'}{(t-t')^2}\left\{
n_0(x')\frac{\partial f_0(p)}{\partial
p}|_{p=m(x-x')/(t-t')}- \right.\nonumber\\
\left. \frac{t}{m}f_0(p)|_{p=m(x-x')/(t-t')} \nabla_x
n_0[x'(1+\frac{t'}{t})-x\frac{t'}{t}] \right\}. \label{F41}
\end{eqnarray}

Therefore, the current $j_1 (x,t)$ for large $t$ takes the form
\begin{eqnarray}
j_1 (x,t)=-e \int_{-\infty}^t dt'exp\,[-\varepsilon (t-t')]\int d
x'\int dp v \left [n_0(x')\frac{\partial f_0(p)}{\partial p}+
\frac{t'}{m} [\nabla_{x'}
n_0(x')] f_0(p)\right] \times \nonumber\\
\delta(x-x'-v(t-t'))E(x',t'). \label{F42}
\end{eqnarray}

Then we arrive at the approximate expression of the "hydrodynamic"
electrical flux in the collisionless case:
\begin{eqnarray}
j_1 (x,t)=-e \int_{t_0}^t dt'exp\,[-\varepsilon (t-t')]\int d x'\int dp v  \times \nonumber\\
 \left[n_0(x') \mu'(x-x',t-t')+\frac{t'}{m} \nabla_{x'}
n_0(x')\mu''(x-x',t-t')\right]E(x',t'),\label{F43}
\end{eqnarray}

where the generalized mobilities are given by
\begin{eqnarray}
\mu' (x,t)= - \int dp v \frac{\partial f_0(p)}{\partial p}
\delta(x-vt) \label{F44}
\end{eqnarray}
and
\begin{eqnarray}
\mu'' (x,t)= - \int dp v f_0(p) \delta(x-vt). \label{F45}
\end{eqnarray}
We can also introduce the mobility operator $\tilde \mu$
\begin{eqnarray}
j_1 (x,t)=e \int_{-\infty}^t dt'\int d x' exp\,[-\varepsilon
(t-t')]E(x',t') \tilde \mu(x,x',t,t') n_0(x'),\label{F46}
\end{eqnarray}
where $\tilde \mu(x,x',t,t')$ equals
\begin{eqnarray}
\tilde \mu(x,x',t,t')=-\int dp v \delta(x-x'-v(t-t'))
 \left[\frac{\partial f_0(p)}{\partial p}+ f_0(p) \frac{t'}{m}
 \nabla_{x'}
\right]. \label{F47}
\end{eqnarray}

Therefore, equation (\ref{F32a}) for the flux perturbation
associated with the presence of the weak electrical field in the
collisionless limit has the form
\begin{eqnarray}
\frac{\partial n_1(x,t)}{\partial t}+ e \nabla_x  \int_{-\infty}^t
dt'\int d x' exp\,[-\varepsilon (t-t')]E(x',t') \int dp v
 \tilde \mu(x,x',t,t') n_0(x')=0. \label{F47a}
\end{eqnarray}

If the space dispersion is negligible $\tilde \mu(x,x',t,t')\sim
\delta(x-x')$ and (\ref{F47a}) transforms into
\begin{eqnarray}
\frac{\partial n_1(x,t)}{\partial t}+ e \int_{-\infty}^t dt'
exp\,[-\varepsilon (t-t')] \tilde \mu(t,t')\nabla_x \left[E(x,t')
  n_0(x)\right]=0. \label{F47b}
\end{eqnarray}
Finally, for the case of slow changing in space of the density
profile $n_0(x)$, when the parameter $\tau_0 <v>/L\ll 1$ ($<v>$ ,
$\tau_0$ and $L$ are the average velocity of the particles, the
characteristic time scale for the electric field and the
characteristic space scale for the density $n_0(x)$ respectively)
the second term in brackets Eq.~(\ref{F47}) can be omitted and the
operator $\tilde \mu$ modifies to the function (\ref{F44})
$\mu'(x-x',t-t')$:
\begin{eqnarray}
\tilde \mu(x-x',t-t')\rightarrow \mu'(x-x',t-t')=-\int dp
v\delta(x-x'-v(t-t'))
 \frac{\partial f_0(p)}{\partial p}. \label{F48}
\end{eqnarray}
Then the diffusion equation (\ref{F47a}) simplifies to the form
typical for the case with an electric field present:
\begin{eqnarray}
\frac{\partial n_1(x,t)}{\partial t}+ e \nabla_x  \int_{-\infty}^t
dt'\int d x' exp\,[-\varepsilon (t-t')]E(x',t') \mu' (x-x',t-t')
n_0(x')=0, \label{F48a}
\end{eqnarray}
Evidently the function $\mu'(x,t)$ is simply connected with the
conductivity $\sigma(x,t)$ (in the case considered  with the
collisionless conductivity) by the equality $\sigma(x,t)=e^2
n_0(x)\mu'(x,t)$.

This consideration provides the evident answer on how the
time-dependent electrical field should be included in the
diffusion equation and permits to make the choice between the
different forms of the diffusion equations considered earlier
[14]. The structure of Eqs.~(\ref{A4}), (\ref{A30}) and
(\ref{F48a}) confirms the result of the generalized diffusion
equation, introduced in the papers [3,4] (on the example of some
particular form of the kernel in the kinetic approximation
considered above).

\section{Stop-move collisions}

Now let us consider on the kinetic level the problem of transport
for the particles, which can move in a time-dependent external
electric filed as the quasi-free particles, but can be trapped and
stay in the rest state during some time. The similar problem has
been consider for the time-independent external field on the basis
of generalized Fokker-Planck equation in [23].

Let us introduce a "collision" integral $I$, that takes into
account the specific "jumps" of the particles:
\begin{eqnarray}
I = - \nu f(p,x,t)+\nu \int_{t_0}^t dt'\psi(t-t') f(p,x,t').
\label{F49}
\end{eqnarray}
Therefore the kinetic equation reads
\begin{eqnarray}
\frac{\partial f(p,x,t)}{\partial t}+v \frac{\partial
f(p,x,t)}{\partial x}+ e E(x,t)\frac{\partial f(p,x,t)}{\partial
p}= \nonumber \\
 - \nu f(p,x,t)+\nu \int_{t_0}^t dt'\psi(t-t') f(p,x,t').
 \label{F50}
\end{eqnarray}

This "stop-move" collision integral describes the moving
particles, which may change from a "moving" state to the "rest"
state and vise versa. We assume that the change from the "rest"
state to "moving" state takes place with recovering of the
momentum distribution. The momentum distribution of the moving
particles, which leave the phase volume $\{dx, dp\}$ at the moment
$t'$ at the point of the phase space $x,p$ is equivalent to the
momentum distribution of the particles, which arises from the
"rest" state at the position $x$ for $t>t'$, with the delay time
$t-t'$.  More complicated situations will be considered in a
separate study. The function $\psi(t)$ characterizes the
probability for the particles to stay in a state of rest during a
time span $t-t'$.

Let us consider the conservation laws for the kinetic equation
with such jumps. The continuity equation reads
\begin{eqnarray}
\frac{\partial n_f(x,t)}{\partial t}+ div{\bf j(x,t)}\equiv \int
dp I(p,x,t) = - \nu n_f(x,t)+ \nu \int_{t_0}^t
dt'\psi(t-t')n_f(x,t'). \label{F51}
\end{eqnarray}

We have distinguished between the "flying" particles and the
particles at "rest" state. The function $f(p,x,t)$ is the
distribution of the "flying" particles ($p\neq 0$). We also
introduce the density of the "rest" ($p=0$) particles $n_r(x,t)$.
We use the "stop-move collision" term for the process of
transferring between the "flying" and the "rest" states.

The conservation of the total number of particles reads
\begin{eqnarray}
\int dx [n_f(x,t)+n_r(x,t)]=N, \nonumber \\
N \equiv N_f+N_r,\label{F52}
\end{eqnarray}
where $N$ is the constant. There is also the evident equality
\begin{eqnarray}
\frac{\partial n_r(x,t)}{\partial t}= \nu n_f(x,t)- \nu
\int_{t_0}^t dt'\psi(t-t')n_f(x,t'). \label{F53}
\end{eqnarray}

From  Eqs.~(\ref{F50}),(\ref{F53}) it follows that
\begin{eqnarray}
\frac{\partial n_r(x,t)}{\partial t}+\frac{\partial
n_f(x,t)}{\partial t}+div{\bf j(x,t)}=0. \label{F54}
\end{eqnarray}
Equations for the numbers of "free" and "rest" particles are
\begin{eqnarray}
\frac{\partial N_f(t)}{\partial t}= - \nu N_f(t)+ \nu \int_{t_0}^t
dt'\psi(t-t')N_f(t'), \label{F55}
\end{eqnarray}
\begin{eqnarray}
\frac{\partial N_r(t)}{\partial t}= \nu N_f(t)- \nu \int_{t_0}^t
dt'\psi(t-t')N_f(x,t'). \label{F56}
\end{eqnarray}
Integration of Eq.~(\ref{F54}) by $x$ leads to Eq.~(\ref{F52}).

Now let us integrate the kinetic equation by $p$ with the
multiplier $p$. The relevant equation of motion reads (dimension
$s=1$)
\begin{eqnarray}
\frac{\partial j(x,t)}{\partial t}+\int dp v^2 \frac{\partial
f(p,x,t)}{\partial x}-\frac {e E(x,t)}{m}
n_f(x,t)=\nonumber \\
- \nu j(x,t)+ \nu \int_{t_0}^t dt'\psi(t-t')j(x,t').  \label{F57}
\end{eqnarray}
We will assume that the integral term with $f(p,x,t)$ in
Eq.~(\ref{F57}) can be represented as $d(t) \,\partial
n_f(x,t)/\partial x$. This the exact representation is exact for,
e.g., such a form of the distribution function $f(p,x,t)=\tilde
f(p,t) n_f(x,t)$. The function $d(t)$ in this case equals:
\begin{eqnarray}
d(t) = \equiv \int dp v^2 \tilde f(p,t) \label{F58}
\end{eqnarray}
For the Maxwellian distribution $d(t)$ is time independent
$d(t)=d=T/m$, where $T$ is the temperature. In general
$d(t)=<v^2>$ is the average velocity of the "flying" particles.

 Then Eq.~(\ref{F57}) represents the integro-differential
connection of $j(x,t)$ and $n_f(x,t)$:
\begin{eqnarray}
\frac{\partial j(x,t)}{\partial t}+ d(t) \frac{\partial
n_f(x,t)}{\partial x}-\frac{e E(x,t)}{m}
n_f(x,t)=\nonumber \\
- \nu j(x,t)+ \nu \int_{t_0}^t dt'\psi(t-t')j(x,t').  \label{F59}
\end{eqnarray}
In order to solve this equation we use the adiabatic switched
process for "hopping collisions" ($t_0=-\infty$) and the
Fourier-transform of Eq.~(\ref{F59}) by time:
\begin{eqnarray}
\left\{-i\omega + \nu[1 - \psi(\omega)]\right\}j(x,\omega) =
\varphi(x,\omega), \label{F60}
\end{eqnarray}
where
\begin{eqnarray}
\psi(\omega)=\int_0^\infty d\tau exp(i\omega\tau) \psi(\tau),
\label{F61}
\end{eqnarray}
and we denote
\begin{eqnarray}
\varphi(x,t)=-d(t) \frac{\partial n_f(x,t)}{\partial x}+\frac{e
E(x,t)}{m} n_f(x,t). \label{F62}
\end{eqnarray}
The solution for the flux is then
\begin{eqnarray}
j(x,t)=\int \frac{d\omega}{2\pi} \frac{exp(-i\omega
t)}{-i\omega+\nu  [1-\psi(\omega)]}\, \varphi(x,\omega)
\label{F63}
\end{eqnarray}
or
\begin{eqnarray}
j(x,t)=\int dt' \int \frac{d\omega}{2\pi} \frac{exp([-i\omega
(t-t')]}{i\omega-\nu  [1-\psi(\omega)]}\left[d(t') \frac{\partial
n_f(x,t')}{\partial x}-\frac{e E(x,t')}{m} n_f(x,t')\right].
 \label{F64}
\end{eqnarray}
The flux can be rewritten by introducing the function $\chi(t-t')$
\begin{eqnarray}
j(x,t)=\int dt' \chi(t-t')\left[d(t') \frac{\partial
n_f(x,t')}{\partial x}-\frac{e E(x,t')}{m} n_f(x,t')\right],
 \label{F65}
\end{eqnarray}
where
\begin{eqnarray}
\chi(t-t')\equiv  \int \frac{d\omega}{2\pi i} \frac{exp[-i\omega
(t-t')]}{\omega+i\nu [1-\psi(\omega)]}. \label{F66}
\end{eqnarray}
Inserting this flux into the continuity equation we find the
diffusion equation in the form
\begin{eqnarray}
\frac{\partial n_f(x,t)}{\partial t}=\int dt'
\chi(t-t')\left\{d(t')  \triangle n_f(x,t')-\frac{e}{m}\nabla
[E(x,t')n_f(x,t')]\right\}, \label{F67}
\end{eqnarray}
which, for time-independent $d$, is the particular case of
Eq.~(\ref{A4}), based on the general master equation for
diffusion, introduced in [3,4]. An essential feature of the
diffusion process is the character of the influence of the
time-dependent external field placed in Eq.~(\ref{F67}) under the
time integral. This equation coincides formally with the
hydrodynamic equation Eq.~(\ref{A4}) if $\chi(t-t')$ is the
retarded function ($\chi(t-t')=0$ for $t<t'$).

\section{Conclusions}

We show that the generalized master equation with two times, which
has been introduced in [1,2] and [14], can describe the influence
of inhomogeneous and time-dependent external fields on the
diffusion processes. Linearization of the general master equation
in the external field leads to essential simplifications. In this
case the diffusion processes depend, in general, on two different
functions of time, which describe retardation, or
frequency-dependent mobility and diffusion, in particular, due to
the finite time of occupation and transferring particles in space
in the presence of the external field. Relations with simpler
models are established. The rigorous consideration on the basis of
the hydrodynamic approach and various kinetic equations confirms
the results of the phenomenological approach of the generalized
master equation. Of cause, the kernel functions $W$ or $P$ can
only be defined in a concrete way in the framework of particular
physical models, e.g., on the basis of kinetic theory with
specific collision integrals, describing the stochastic motion
with retardation. We also introduced the new stop-move collision
integral, which describes the processes of diffusion with
particles continuously changing from moving to resting and back.
The appropriate kinetic equation is solved for a time-dependent
external field, which also confirms the results of the diffusion
master equation approach. This type of motion is very common in
Nature and the introduced collision integral can easily be
generalized to more complex processes of "stop-move" motion. The
analysis presented in this paper opens opportunities to consider a
wide class of the problems of normal and anomalous transport in
external fields on the basis of the generalized master equation
with two times. The Einstein relations in general are not
applicable to the case of the non-stationary external field, but
in the particular case can be valid for the time dependent
diffusion and mobility functions, as it was found above in the
present paper.

\section*{Acknowledgment}
The authors are thankful to W. Ebeling, A.M. Ignatov and Yu.P.
Vlasov for valuable discussions of the problems, reflected in this
paper. This work has been supported by The Netherlands
Organization for Scientific Research (NWO) and the Russian
Foundation for Basic Research.


\begin{thebibliography}{99}
\bibitem{1}
S.A. Trigger, G.J.F. van Heijst, P.P.J.M. Schram, \emph{Physica
A}, 347, 77 (2005); http://arXiv.org/abs/physics/0401327
\bibitem{2} S.A. Trigger, G.J.F. van Heijst, P.P.J.M. Schram,
\emph{J.of Physics: Conference Series}, 11, 37 (2005)

\bibitem{3}
E.W. Montroll and M.F. Schlezinger, in \emph{Studies in
Statistical Mechanics}, edited by J. Leibowitz and E.W. Montroll,
V. 2 (North-Holland, Amsterdam, 1984)
\bibitem{4}
R. Mantegna, H. Stanley, \emph{Nature} 376, 46 (1995)
\bibitem{5}
P. Gopikrishnan, V. Plerou, L. A. N. Amaral, M. Meyer, H.E.
Stanley, \emph{Phys. Rev.} E60, 5305 (1999).

\bibitem{6}
M. Newman, \emph{Contemporary Physics} 46, 323 (2005)
\bibitem{7}
P.K. Schukla, A.A. Mamun, \emph{Introduction to Dusty Plasma
Physics}, Institute of Physics Publishing, Bristol, 2002.
\bibitem{8}
W.H. Stacey, \emph{Nuclear reactor physics}, Wiley Publishing,
2001.
\bibitem{9}
Lecture Notes in Physics, LNP 557, \emph{Stochastic Processes in
Physics, Chemistry and Biology}, Edited by J.A. Freund and T.
Poschel, Springer, 2002.
\bibitem{10}
A.S. Monin and A.M. Yaglom, \emph{Statistical Fluid Mechanics:
Mechanics of Turbulence }(MIT Press, Cambridge, MA, 1975), Vol II.

\bibitem{11}
B. Rinn, P. Mass, J.P. Bouchaud, \emph{Phys. Rev. Lett.} 84, 5405
(2000).
\bibitem{12}
B.J. West, M. Bologna, P. Grigolini, {\it Physics of Fractal
Operators}, (Springer-Verlag New York), 2003.
\bibitem{13}
I.M. Sokolov, \emph{Phys. Rev.} E73, 067102 (2006) and I. Sokolov
and J. Klafter, \emph {Phys. Rev. Lett.} 97, 140602 (2006).

\bibitem{14} S.A. Trigger, Genearlized Master Equation with Two Times:
Diffusion in External Field, \emph{Phys. Letters A}, in print;
http://arXiv: cond-mat/0608060, July 2006.

\bibitem{15}
V.M. Kenkre, E. Montroll, M.F. Shlesinger, \emph{J. Stat. Phys.}
9, 45 (1973)
\bibitem{16}
H. Scher and E. Montroll, \emph{Phys. Rev.} B12, 2455 (1975)
\bibitem{17}
V.M. Kenkre and R.S. Knox \emph{Phys. Rev.}  B 9, 5279, (1974)
\bibitem{18}
J. Klafter, A. Blumen, M.F. Shlesinger, \emph{Phys. Rev.} A, 35,
3081 (1987)
\bibitem{19}
V.Yu. Zaburdaev, \emph {J. Stat. Phys.} , 123, pp. 871-881, (2006)

\bibitem{20}
R. Metzler, E. Barkai, J. Klafter, \emph{Phys. Rev. Lett.}, 82,
3563 (1999)
\bibitem{21}
R. Metzler, E. Barkai, J. Klafter, \emph{Europhys. Lett.}, 46, 431
(1999)
\bibitem{22}
W. Ebeling, \emph{Contr. Plasma Phys.} 7, 11 (1967)
\bibitem{23}
R. Metzler, J. Klafter, \emph{Phys. Rev.} E, 61, 6308 (2000)



\end{thebibliography}
\end{document}